\documentclass[twocolumn,amsmath,amssymb,10pt,superscriptaddress,a4paper,letterpaper,fleqn]{revtex4-1}
\usepackage{amssymb}
\usepackage{epsfig}
\usepackage{graphicx}
\usepackage{dcolumn}
\usepackage{array}
\usepackage{bm}
\usepackage{fancyheadings}
\usepackage{longtable}
\usepackage{multirow}
\usepackage{float}
\usepackage{afterpage}
\usepackage{color}

\setlongtables
\usepackage[breaklinks=true,linkbordercolor={1 1 1}]{hyperref}

\begin{document}
\setcounter{page}{1}


\title{
\qquad \\ \qquad \\ \qquad \\  \qquad \\  \qquad \\ \qquad \\
Experimental Neutron-Induced Fission Fragment Mass Yields of $^{232}$Th and $^{238}$U at Energies
from 10 to 33 MeV}

\author{V.D. Simutkin}
\email[Corresponding author: ]{vasily.simutkin@physics.uu.se}
\affiliation{Division of Applied Nuclear Physics, Department of Physics and Astronomy, Uppsala University, Box 516, SE-751 20 Uppsala, Sweden}

\author{S. Pomp}
\affiliation{Division of Applied Nuclear Physics, Department of Physics and Astronomy, Uppsala University, Box 516, SE-751 20 Uppsala, Sweden}

\author{J. Blomgren} 
\affiliation{Division of Applied Nuclear Physics, Department of Physics and Astronomy, Uppsala University, Box 516, SE-751 20 Uppsala, Sweden} 

\author{M. \"Osterlund} 
\affiliation{Division of Applied Nuclear Physics, Department of Physics and Astronomy, Uppsala University, Box 516, SE-751 20 Uppsala, Sweden} 

\author{R. Bevilacqua} 
\affiliation{Division of Applied Nuclear Physics, Department of Physics and Astronomy, Uppsala University, Box 516, SE-751 20 Uppsala, Sweden} 

\author{P. Andersson} 
\affiliation{Division of Applied Nuclear Physics, Department of Physics and Astronomy, Uppsala University, Box 516, SE-751 20 Uppsala, Sweden} 

\author{I.V. Ryzhov} 
\affiliation{Khlopin Radium Institute, 2nd Murinski pr. 28, 194021, St. Petersburg, Russia} 

\author{G.A. Tutin} 
\affiliation{Khlopin Radium Institute, 2nd Murinski pr. 28, 194021, St. Petersburg, Russia} 

\author{S.G. Yavshits} 
\affiliation{Khlopin Radium Institute, 2nd Murinski pr. 28, 194021, St. Petersburg, Russia} 

\author{L.A. Vaishnene} 
\affiliation{Petersburg Nuclear Physics Institute of Russian Academy of Science, 188350, Gatchina, Leningrad district, Russia} 

\author{M.S. Onegin} 
\affiliation{Petersburg Nuclear Physics Institute of Russian Academy of Science, 188350, Gatchina, Leningrad district, Russia} 

\author{J.P. Meulders} 
\affiliation{FNRS and Institute of Nuclear Physics, Universit\' e catholique de Louvain, B-1348 Louvain-la-Neuve, Belgium} 

\author{R. Prieels} 
\affiliation{FNRS and Institute of Nuclear Physics, Universit\' e catholique de Louvain, B-1348 Louvain-la-Neuve, Belgium} 

\date{\today} 

\begin{abstract}
Development of nuclear energy applications requires data for neutron-induced reactions for actinides in a wide neutron energy range. Here we describe measurements of pre-neutron emission fission fragment mass yields of $^{232}$Th and $^{238}$U at incident neutron energies from 10 to 33 MeV. The measurements were done at the quasi-monoenergetic neutron beam of the Louvain-la-Neuve cyclotron facility CYCLONE; a multi-section twin Frisch-gridded ionization chamber was used to detect fission fragments. For the peak neutron energies at 33, 45 and 60 MeV, the details of the data analysis and the experimental results have been published before and in this work we present data analysis in the low-energy tail of the neutron energy spectra. The preliminary measurement results are compared with available experimental data and theoretical predictions.
\end{abstract}
\maketitle

\lhead{ND 2013 Article $\dots$}
\chead{NUCLEAR DATA SHEETS}
\rhead{A. Author1 \textit{et al.}}
\lfoot{}
\rfoot{}
\renewcommand{\footrulewidth}{0.4pt}

\section{ INTRODUCTION}
Emerging nuclear energy applications and nuclear theory require nuclear data on neutron-induced reactions at energies higher than thermal. However, the data on fission fragment mass yields are scarce at neutron energies above 10 MeV. We have measured neutron-induced fission fragment mass yields of $^{232}$Th and $^{238}$U at the Louvain-la-Neuve neutron beam facility at quasi-monoenergetic neutron beams with the peak energies 33, 45, and 60 MeV. The incident neutron energy spectrum consisted of a high-energy peak produced by the $^7$Li(p,n)$^7$Be reaction and a low-energy tail containing about 50\% of all produced neutrons. The measurements results for the peak energies have been published in the Ref.~\cite{11Ryz}; here we present results from 10 to 33 MeV extracted from the low-energy tail.
\section {INSTRUMENT AND METHOD}
Comprehensive information about the experiment can be found in Ref.~\cite{11Ryz} so only a brief overview is given here. A multi-section Frisch-gridded ionization chamber has been used to detect fission fragments and to measure their kinetic energies. The double kinetic energy (2E) method was used to extract fission fragment mass yields. To determine the energy of a non-peak incident neutrons induced fission, we used the time-of-flight (TOF) difference between them and the peak neutrons.
Due to a periodic structure of the cyclotron proton beam, experimental fission fragment mass yields suffer from the contribution of low-energy neutrons (up to 10\%) from the previous proton bursts.  Wrap-around neutron-induced fission fragment mass distributions were calculated with the PYF code(Ref.~\cite{08Gor}) at energies above 5 MeV and with the GEF code(Ref.~\cite{10Sch}) at energies below 5 MeV.
\section {RESULTS}
\subsection{EXPERIMENTAL NEUTRON-INDUCED FISSION FRAGMENT MASS YIELDS}
\begin{figure}[!ht]
\includegraphics[width=0.95\columnwidth]{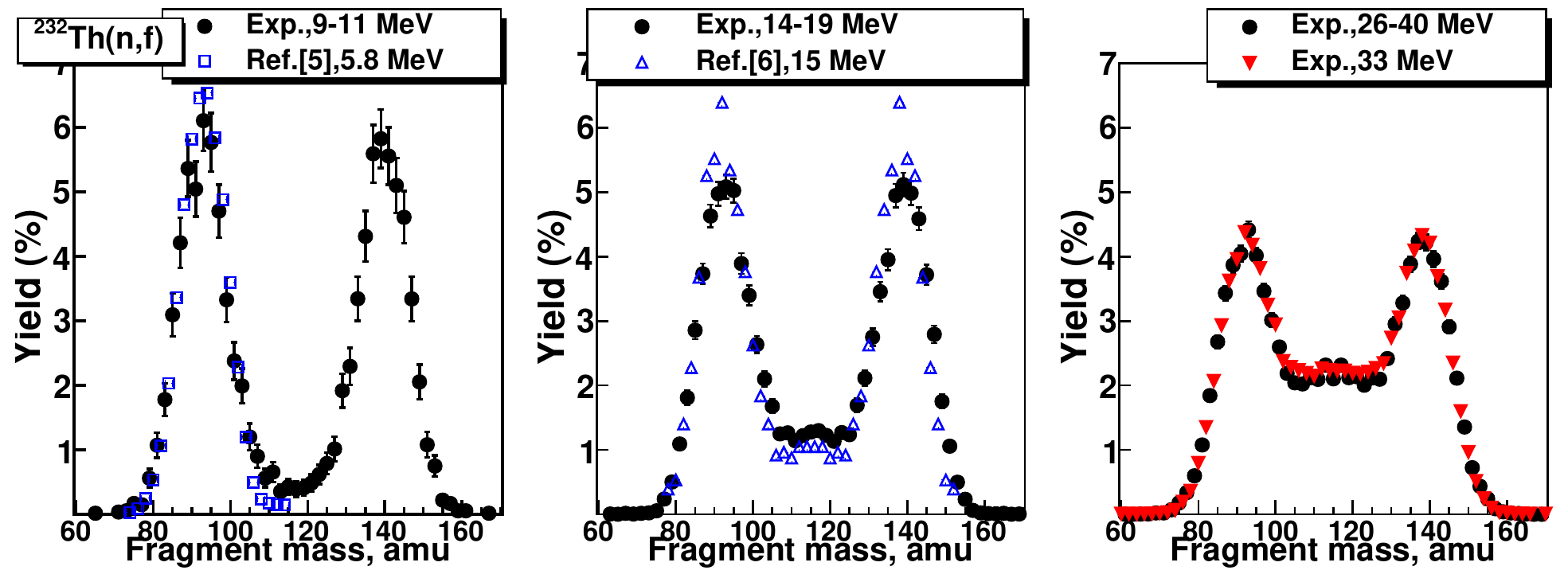}
\caption{\label{fig:Th_Yield}Experimental pre-neutron emission fission fragment mass yields of $^{232}$Th for the incident neutron energy intervals 9-11 MeV, 14-19 MeV, and 26-40 MeV (from left to right). The measured results are shown with full symbols, and available experimental data with open symbols. The result for the interval 26-40 MeV is also compared with the measurement result for the high-energy peak at 33 MeV.}
\end{figure}

\begin{figure}[!ht]
\centering
\includegraphics[width=0.95\columnwidth]{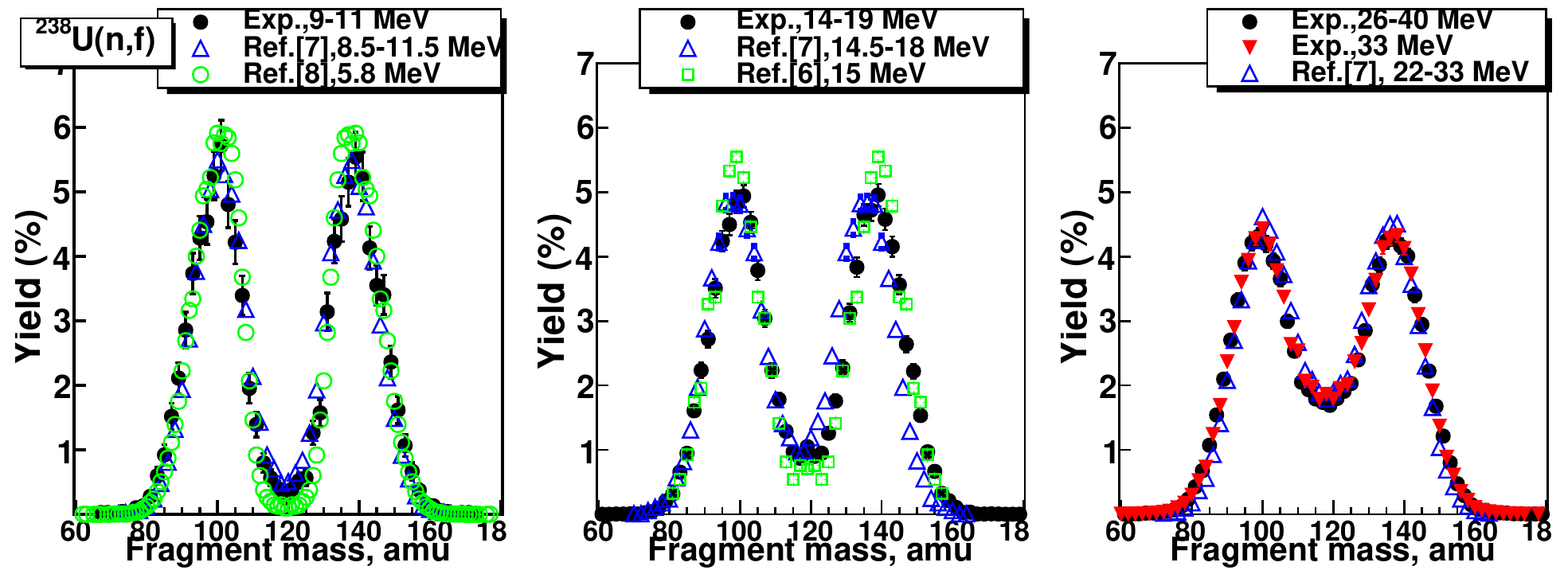}
\caption{\label{fig:U_Yield}Experimental pre-neutron emission fission fragment mass yields of $^{238}$U for the incident neutron energy intervals 9-11 MeV, 14-19 MeV, and 26-40 MeV (from left to right). The measured results are shown with full symbols, and available experimental data with open symbols. The result for the interval 26-40 MeV is also compared with the measurement result for the high-energy peak at 33 MeV.}
\end{figure}

To select the intervals of incident neutron energy, we took into consideration the uncertainty $\sigma_{tof}$=3 ns of the incident neutron time-of-flight measurement as well as the incident neutron energy upper and lower limits set by the peak energy of the incident neutron energy spectrum and the cyclotron RF respectively. Incident neutron energy intervals 9-11 MeV, 11-14 MeV, 14-19 MeV, 19-26 MeV, and 26-40 MeV were used for the data analysis. The interval 26-40 MeV with the average energy 33 MeV has also been used for the consistency check of our data analysis procedure by comparison with the high-energy peak data at 33 MeV. The measurement results for intervals 9-11 MeV, 14-19 MeV, and 26-40 MeV are shown in Figs.~\ref{fig:Th_Yield} and~\ref{fig:U_Yield}.
\subsection{SYMMETRIC FISSION PROBABILITY}
\begin{figure}[!ht]
\centering
\includegraphics[width=0.95\columnwidth]{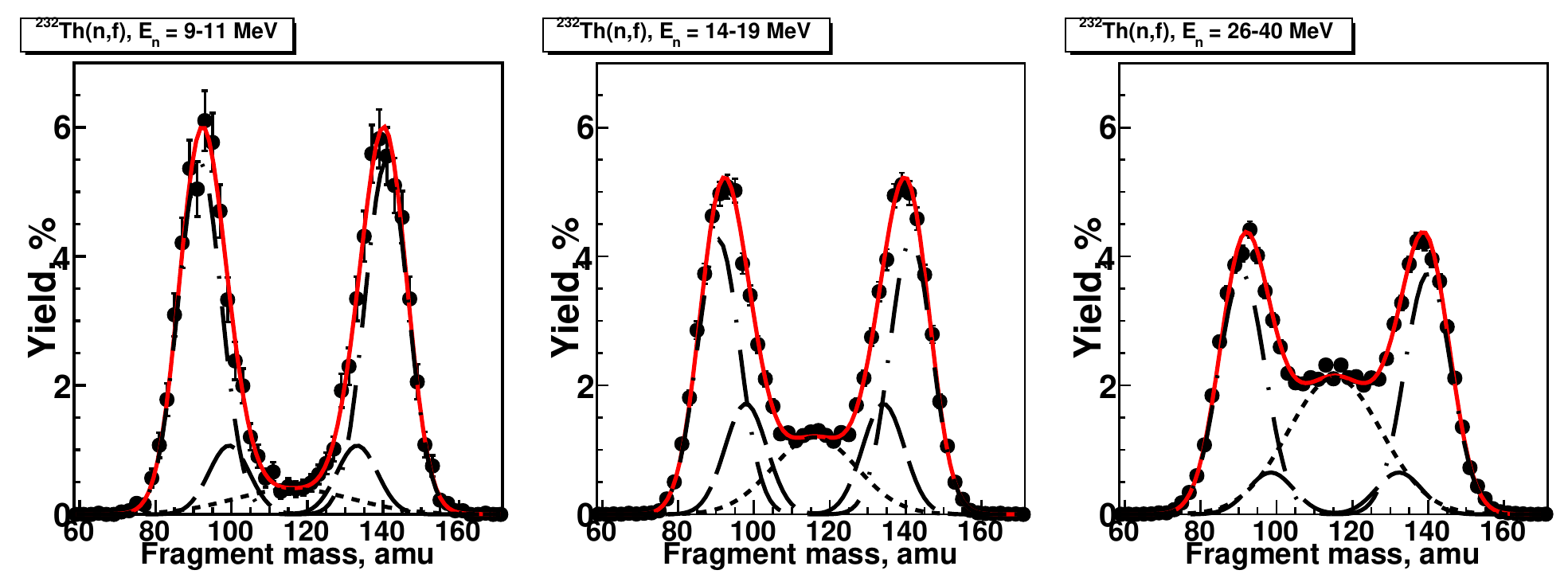}
\caption{Fitted experimental pre-neutron emission fission fragment mass yields of $^{232}$Th for the incident neutron energy intervals 9-11, 14-19, and 26-40 MeV (from left to right). Symbols are used for the experimental results, and solid line for the fit. The contributions of different modes are shown with the dash-dash line for the STI mode, the dash-dot line for the STII mode, and the dot-dot line for the SL mode.}
\label{Th_Fit}
\end{figure}

\begin{figure}[!ht]
\centering
\includegraphics[width=0.95\columnwidth]{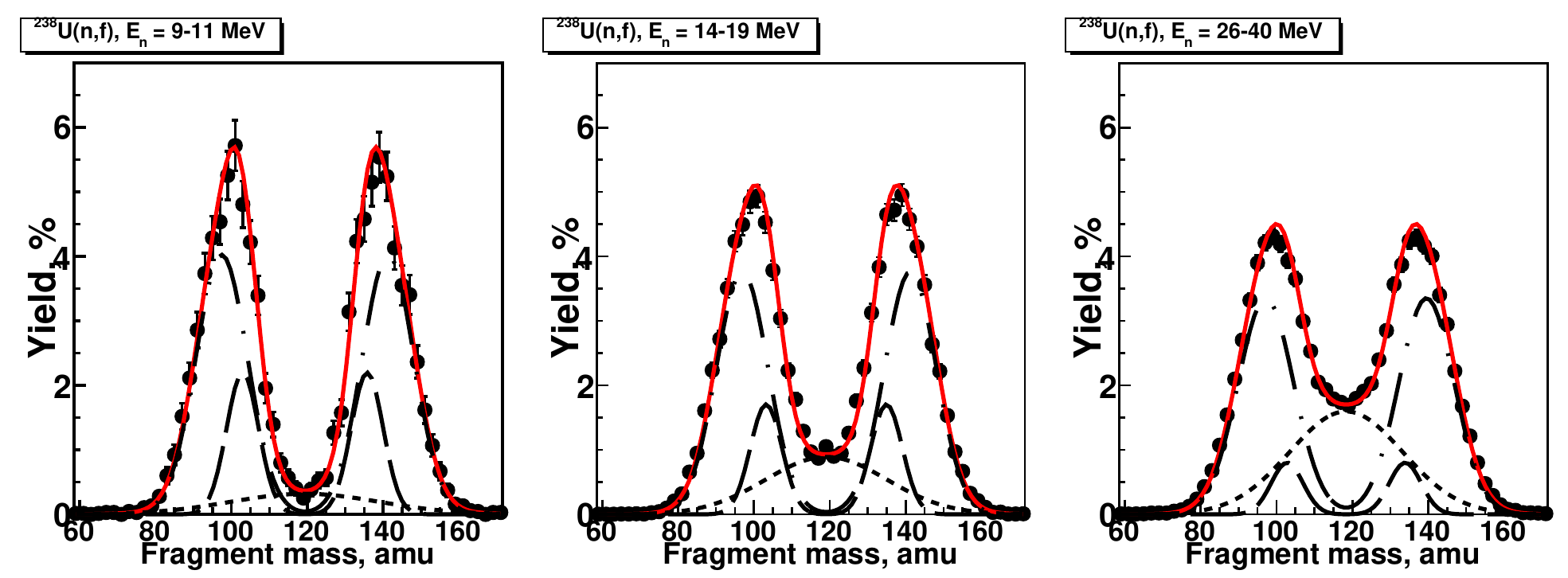}
\caption{Fitted experimental pre-neutron emission fission fragment mass yields of $^{238}$U for the incident neutron energy intervals 9-11, 14-19, and 26-40 MeV (from left to right). Symbols are used for the experimental results, and solid line for the fit. The contributions of different modes are shown with the dash-dash line for the STI mode, the dash-dot line for the STII mode, and the dot-dot line for the SL mode.}
\label{U_Fit}
\end{figure}

To parametrize the experimental results, we fitted experimental total kinetic energy-fragment (TKE) vice mass (A) yields Y(A,TKE) with a model function as in Ref.~\cite{98Obe}. The fitting results projected on the fragment mass plane are shown in Figs.~\ref{Th_Fit} and~\ref{U_Fit} in comparison with the measured results.

\begin{figure}[!ht]
\centering
\includegraphics[width=0.95\columnwidth]{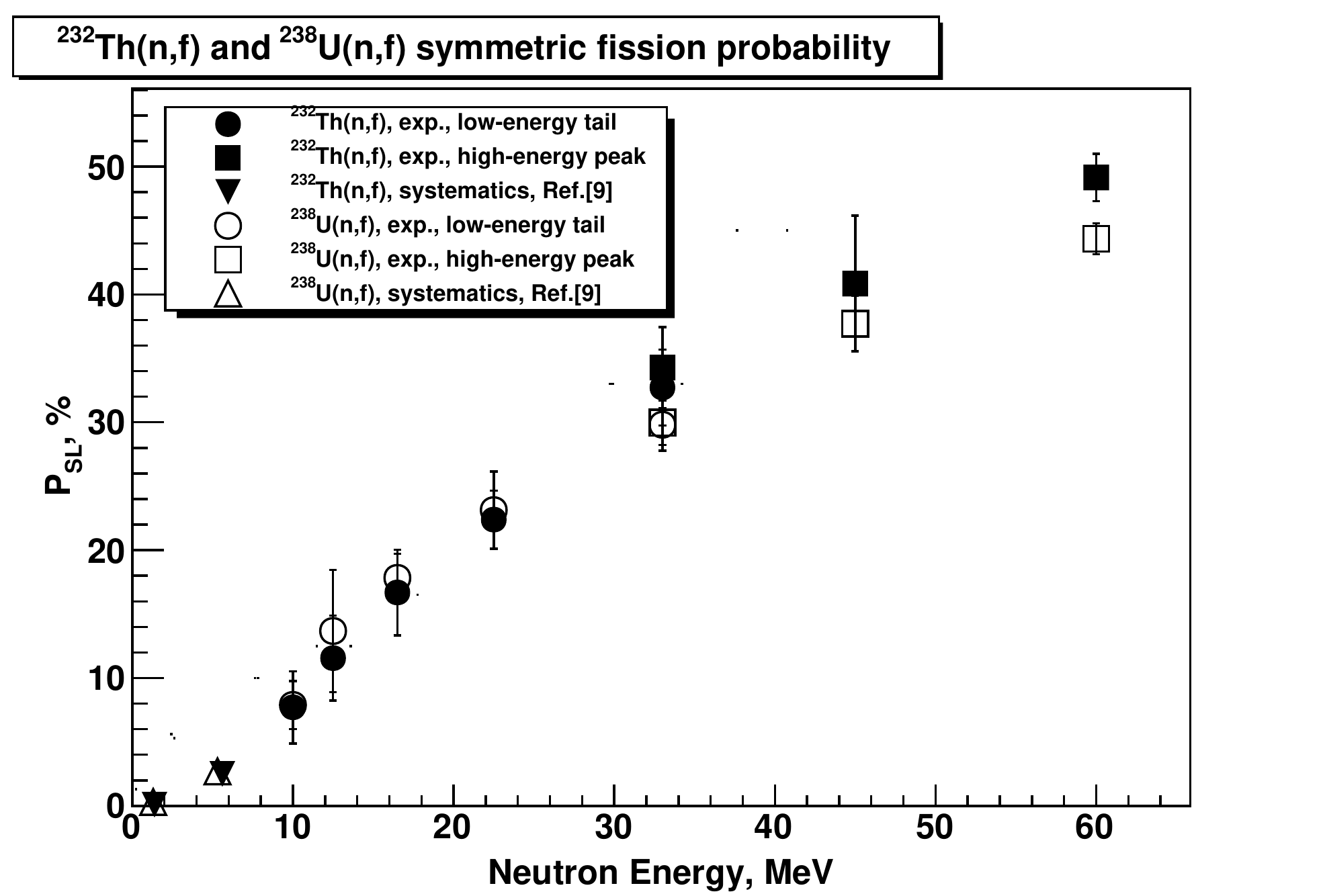}
\caption{Symmetric fission probability P$_{SL}$ of $^{232}$Th and $^{238}$U as a function of the incident neutron energy. The average energy is used for the incident neutron energy intervals. Results for $^{232}$Th are shown with open symbols, and results for $^{238}$U with full symbols. Circles are used for the low-energy tail data, squares for the high-energy peak data, and triangles for the systematics from Ref.~\cite{99Bro}.}
\label{Psym}
\end{figure}

As one can see from Figs.~\ref{fig:Th_Yield} and~\ref{fig:U_Yield}, symmetric fission probability with the incident neutron energy of $^{232}$Th and $^{238}$U is increased with the incident neutron energy. To quantify this increase, we calculated the contribution of the superlong (SL) mode into Y(A,TKE) yields for each energy interval as well as for high-energy peak neutron energies 33, 45, and 60 MeV. The extracted probability P$_{SL}$ of the SL-mode as a function of the incident neutron energy is shown in Fig~\ref{Psym}.
\subsection{COMPARISON WITH MODEL CALCULATIONS}
\begin{figure}[!ht]
\centering
\includegraphics[width=0.95\columnwidth]{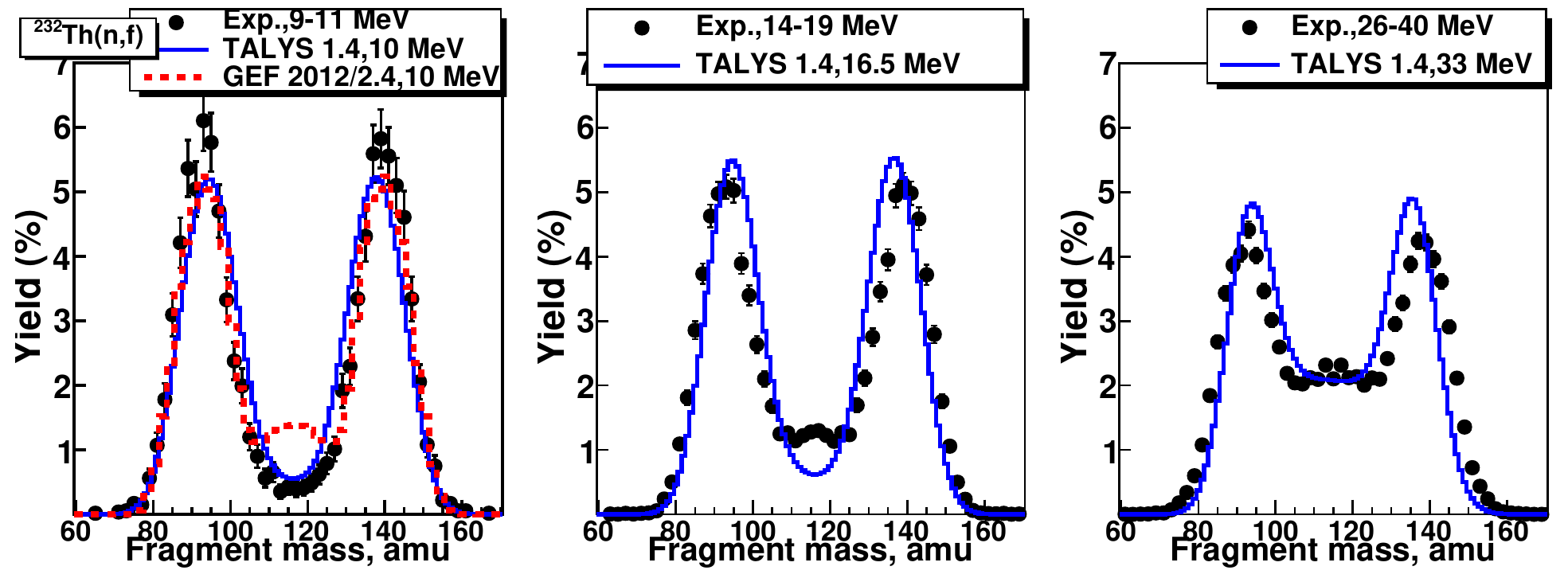}
\caption{Experimental pre-neutron emission fission fragment mass yields of $^{232}$Th for the incident neutron energy intervals 9-11, 14-19, and 26-40 MeV (from left to right) in comparison with the TALYS 1.4 (solid line) and GEF 2012/2.4 (dashed line) calculation results.}
\label{Th_TALYS}
\end{figure}

\begin{figure}[!ht]
\centering
\includegraphics[width=0.95\columnwidth]{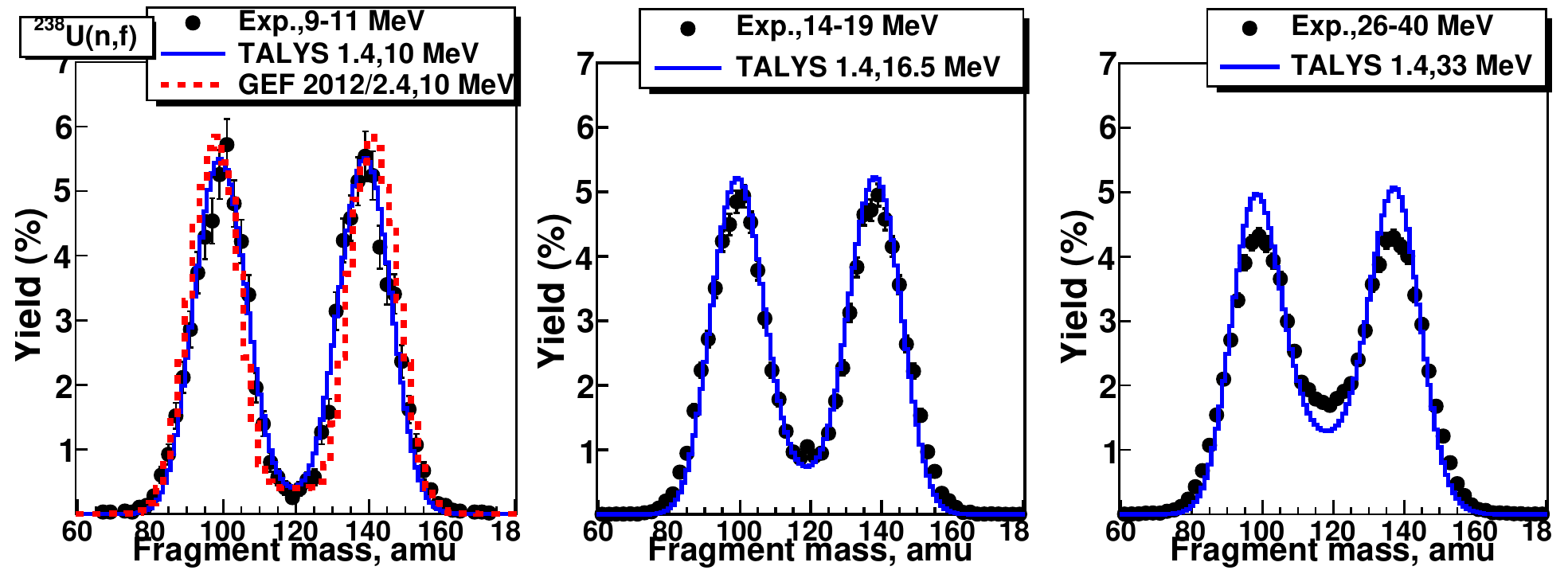}
\caption{Experimental pre-neutron emission fission fragment mass yields of $^{238}$U for the incident neutron energy intervals 9-11, 14-19, and 26-40 MeV (from left to right) in comparison with the TALYS 1.4 (solid line) and GEF 2012/2.4 (dotted line) calculation results.}
\label{U_TALYS}
\end{figure}

We have also compared our experimental results with model calculations done with the nuclear codes TALYS 1.4 (Ref.~\cite{08Kon}) and GEF 2012/2.4. The calculated fragment mass yields are shown in Figs~\ref{Th_TALYS} and~\ref{U_TALYS} as well as the measurement results.
\section{DISCUSSION}
The experimental results agree reasonably well with the available experimental results for both $^{232}$Th and $^{238}$U. However, as we can see in Fig.~\ref{fig:U_Yield}, the compared data for $^{238}$U data from Ref.~\cite{95Zol} have been measured for the energy interval 22-33 MeV and the average energy 27.5 MeV whereas our data for the energy interval 26-40 MeV and the average energy 33 MeV. The seeming agreement indicates that symmetric fission probability values of $^{238}$U are higher in the work~\cite{95Zol} than in our experiment at the same incident neutron energy. This trend was discussed in more detail in Ref.~\cite{11Ryz}.

For the consistency check of the data analysis procedure, we have compared low-energy tail experimental results for the 26-40 MeV energy interval with the high-energy peak data at 33 MeV. As shown in Figs.\ref{fig:Th_Yield} and ~\ref{fig:U_Yield}, the agreement is very good for both $^{232}$Th and $^{238}$U which validates the applicability of the method. The results of the fitting of the Y(A,TKE) yields (Fig.~\ref{Psym}) also confirm that.

Agreement with the nuclear code TALYS 1.4 is good for $^{238}$U for the 9-11 MeV and 14-19 MeV neutron energy intervals and worse for the 26-40 MeV interval. In the case of $^{232}$Th the agreement is satisfactory for the 9-11 MeV and 26-40 MeV intervals. However, our data and the calculated yields for $^{232}$Th systematically disagree in the position of the heavy fragment mass peak. The agreement with the GEF 2012/2.4 code is better for $^{238}$U for the 9-11 MeV interval than in case of $^{232}$Th. The code GEF 2012/2.4, however, does not take into consideration multi-chance fission and is intended for use at energies below 5 MeV.
Also, one has to keep in mind that our data have not been corrected for mass resolution which may modify peak-to-valley ratio of the fragment mass yields.

\section{ACKNOWLEDGEMENT}
This work was supported in part by the International Science and Technology Center (project 3192) and the European Commission within the Sixth Framework Programme through I3-EURONS (contract no. RII3-CT-2004-506065).

\end{document}